\documentclass[letterpaper,twocolumn,10pt]{article}
\usepackage[margin=0.72in,columnsep=0.28in]{geometry}
\usepackage{amsmath,amssymb}
\usepackage{booktabs}
\usepackage{graphicx}
\usepackage{xspace}
\usepackage{multirow}
\usepackage{array}
\usepackage{mathptmx}
\usepackage[T1]{fontenc}
\usepackage[kerning,spacing]{microtype}
\usepackage{flushend}
\usepackage{cite}
\usepackage{xurl}
\usepackage{xcolor}
\usepackage[hidelinks]{hyperref}
\newcommand{\systemname}{PipeAxes}
\newcommand{\papertitle}{\systemname: Decoupling Compute and Communication Granularities for Pipeline-Parallel LLM Serving}
\newcommand{\authoremail}[1]{\href{mailto:#1}{\small\texttt{#1}}}
\hypersetup{
  pdftitle={\papertitle},
  pdfauthor={Xiaoqiang Guo, Yifan Lin, Yang Liu, Jianan Zhang, Yingqi Peng, Jiamei Lv, Yi Gao, Wei Dong},
  pdfkeywords={large language models; pipeline parallelism; inference serving; heterogeneous GPU systems}
}

\newcommand{\sys}{\textup{\textsc{\systemname}}\xspace}
\newcommand{\revisit}{\textup{\textsc{Revisit-PP}}\xspace}
\newcommand{\layerspool}{\texttt{LayerSpool}\xspace}
\newcommand{\streamparent}{\texttt{StreamingParent}\xspace}
\newcommand{\pplink}{\texttt{pp\_link}\xspace}
\newcommand{\hintfour}{\texttt{H-INT4}\xspace}

\makeatletter
\renewcommand\subsection{\@startsection{subsection}{2}{\z@}%
  {-1.9ex\@plus -0.6ex \@minus -0.2ex}%
  {0.9ex \@plus 0.2ex}%
  {\normalfont\normalsize\bfseries}}
\makeatother

\begin{document}
\setlength{\abovedisplayskip}{4pt plus 1pt minus 1pt}
\setlength{\belowdisplayskip}{4pt plus 1pt minus 1pt}

\date{}

\title{\papertitle}

\author{%
  \begin{tabular}{@{}cccc@{}}
    \textbf{Xiaoqiang Guo} & \textbf{Yifan Lin} &
    \textbf{Yang Liu} & \textbf{Jianan Zhang} \\
    \authoremail{guoxq@zju.edu.cn} & \authoremail{015@zju.edu.cn} &
    \authoremail{liu-yang@zju.edu.cn} & \authoremail{zhangjianan@zju.edu.cn} \\[0.6em]
    \textbf{Yingqi Peng} & \textbf{Jiamei Lv} &
    \textbf{Yi Gao} & \textbf{Wei Dong} \\
    \authoremail{yingqipeng@zju.edu.cn} & \authoremail{lvjm@zju.edu.cn} &
    \authoremail{gaoyi@zju.edu.cn} & \authoremail{dongw@zju.edu.cn} \\[0.45em]
    \multicolumn{4}{c}{Zhejiang University, Hangzhou, China}
  \end{tabular}%
}

\maketitle

%-------------------------------------------------------------------------------
\begin{abstract}
%-------------------------------------------------------------------------------
Pipeline parallelism can serve long-context language models on PCIe servers and across heterogeneous GPU clusters. Existing engines split prefill into token chunks to limit decode stalls. The chunk size also sets the GPU batch width and the size of activation transfers between stages. Small chunks reload model weights more often. Large chunks can keep the GPU and the link busy long enough to delay interactive requests.

\sys runs wide prefill batches a few layers at a time so decodes can run between slices. A fused Triton Hadamard-INT4 codec compresses activations, which travel in independently sized row groups over parallel TCP sockets. The scheduler uses latency slack to choose batch width, slice depth, and row-group size. We evaluated \sys in 97 replays covering 5,016 requests on A800, RTX 4090, and RTX 3090 GPUs. All 74 requests met their latency targets during a tenfold bandwidth swing. In a four-stage mixture-of-experts pipeline, \sys served requests within their latency targets at $23.3\times$ the rate of calibrated Revisit-PP and $4.44\times$ the rate of Fixed-4096 with Express. In a mixed 64K-context workload, p95 time to first token for short requests fell from 129.8~s with native vLLM to 512~ms. With INT4 transport, MMLU accuracy rose by 0.06 percentage points and GSM8K accuracy fell by 0.46 percentage points relative to BF16.
\end{abstract}


%-------------------------------------------------------------------------------
\section{Introduction}
\label{sec:intro}
%-------------------------------------------------------------------------------

Serving long-context language models can require several GPUs even when those GPUs share only a PCIe bus or an Ethernet link. Prior systems have explored serving across heterogeneous GPUs and machines~\cite{jiang2024hexgen,griggs2024melange,fang2026planetserve}. We study interactive chat and multi-turn coding agents~\cite{yang2024sweagent} with contexts up to 64K tokens on dense and mixture-of-experts (MoE) models from Qwen3~\cite{qwen3tech2025}. Goodput measures how many requests per second meet their service-level objectives (SLOs)~\cite{agrawal2024sarathi,zhong2024distserve,chen2026muxwise}. We require each request to meet the targets for time to first token (TTFT), time per output token (TPOT), and tail inter-token latency (ITL).

The amount of data that crosses the link depends on how the model is partitioned. Megatron-style tensor parallelism (TP) performs two synchronous collectives per layer~\cite{shoeybi2019megatron}, or $72\text{--}128$ per decode step for the models we study. Prefill--decode disaggregation places the two phases on separate instances~\cite{zhong2024distserve,patel2024splitwise,qin2025mooncake}. On a cache miss, the full prompt KV cache must reach the decode host. This requires $96\text{--}256$~KiB/token for our models, or $103\text{--}275$~s for a 32K prompt at $250$~Mbps before transfer overhead. Pipeline parallelism (PP) partitions layers across sequential stages~\cite{huang2019gpipe,narayanan2019pipedream,hwang2026revisitpp}. Each stage keeps its KV cache local and sends boundary activations to the next stage.

\begin{figure*}[t]
\centering
\includegraphics[width=\textwidth]{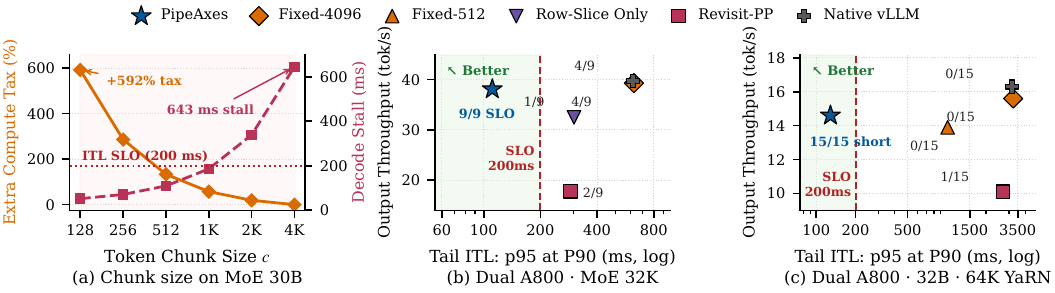}
\caption{\textbf{The cost of small and large prefill chunks.} In panel a, smaller chunks increase stage compute time by up to $591.7\%$ through repeated weight loads on an A800 running the first stage of Qwen3-30B-A3B. Panels b and c show output throughput and tail ITL on two PCIe A800s for 32K MoE and 64K dense workloads. Tail ITL is the p90 across requests of each request's p95 ITL. Labels count all requests that meet their SLOs in panel b and short requests in panel c. The green bands show ITL within the 200~ms target.}
\label{fig:pareto_dilemma}
\end{figure*}

Prefill can block decode on a shared pipeline stage and leave other stages idle~\cite{hwang2026revisitpp}. vLLM~\cite{vllm2026release} and Sarathi-Serve~\cite{agrawal2024sarathi} use chunked prefill. \revisit updates its latency predictor by recursive least squares (RLS) to choose the chunk size and uses delay scheduling to spread out decode work~\cite{hwang2026revisitpp}. When a stage executes and sends one token chunk at a time, the same chunk size sets the GPU batch width, the time until decode can run, and the size of the transfer.

On Qwen3-30B-A3B, the routed experts load nearly all their weights even for narrow chunks. Cutting the chunk size from 4096 to 512 tokens adds $133.6\%$ to stage compute time, and cutting it to 128 adds $591.7\%$. With 4096-token chunks, tail ITL reaches $627.2$~ms on the 32K MoE workload and $3,232.3$~ms on the 64K dense workload in Figure~\ref{fig:pareto_dilemma}. Large chunks also put $64\text{--}80$~MiB BF16 transfers on the cross-node link, and in our tests short prompts wait up to $158.2$~s. Section~\ref{sec:motiv} measures these costs and derives the contribution of repeated weight loads.

\sys keeps prefill batches wide and limits each GPU step to a few layers. The stage can then run decodes between slices while loading each layer's weights once per prefill batch. In Section~\ref{sec:layer}, we show how it reuses KV slot mappings across slices, runs decode on separate CUDA Graphs, and samples the first token as soon as the last prefill slice completes.

A fused Triton codec combines the residual and hidden states, applies a Walsh--Hadamard rotation, and packs the result in INT4. It runs in $0.10$~ms for small batches and reduces the payload by $7.53\times$. The transport in Section~\ref{sec:wire} sends bulk traffic in row groups over parallel TCP sockets so short requests can use the link at frame boundaries. The scheduler in Section~\ref{sec:sched} uses latency slack and measurements of link capacity to choose batch width, slice depth, and row-group size.

We implement \sys in vLLM v0.23.0~\cite{vllm2026release}. Section~\ref{sec:eval} evaluates it on dual PCIe A800s and across RTX 4090 and RTX 3090 nodes. On a four-stage MoE pipeline, it achieves $23.3\times$ the SLO goodput of calibrated \revisit. When link bandwidth changes by a factor of ten, all 74 requests meet their latency targets.

%-------------------------------------------------------------------------------
\section{Background and Motivation}
\label{sec:motiv}
%-------------------------------------------------------------------------------

\subsection{Communication over PCIe and Ethernet}
\label{sec:motiv_pp}

Prior work studies language model serving on PCIe-connected GPUs~\cite{hwang2026revisitpp} and across machines~\cite{borzunov2023petals,jiang2024hexgen,fang2026planetserve}. Section~\ref{sec:eval_setup} describes our PCIe Gen4 testbed and Ethernet links shaped to $100\text{--}1000$~Mbps with up to $30$~ms round-trip time (RTT).

\begin{figure*}[t]
\centering
\includegraphics[width=\textwidth]{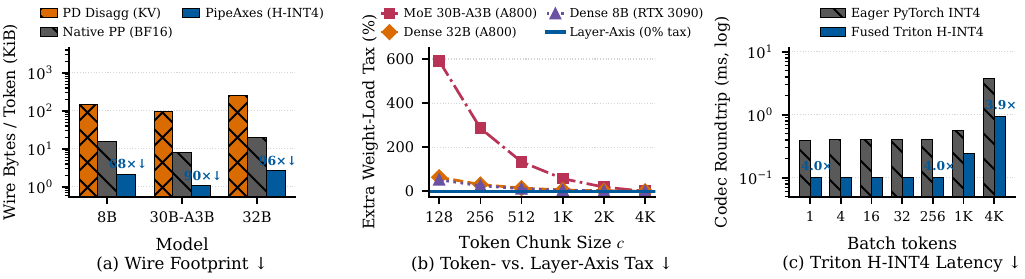}
\caption{\textbf{Transfer size, weight loads, and codec latency.} Panel a shows cross-node bytes per prompt token for the three Qwen3 models. The fused \hintfour codec reduces traffic by $7.53\times$ relative to native PP and by $67.8\text{--}96.4\times$ relative to full KV transfers. Panel b measures the extra compute from repeated weight loads in a 4096-token batch. Layer slicing adds none. Panel c compares codec round-trip times with 4096 hidden channels on an RTX~4090. Panels a and c use log scales.}
\label{fig:characterization}
\end{figure*}

At a $30$~ms RTT, the $2L$ sequential collectives in TP can add $2,160\text{--}3,840$~ms per token for Qwen3 models with 36--64 layers. Communication alone exceeds our $100$~ms TPOT target by $21\text{--}38\times$.

Prefill--decode disaggregation transfers the KV cache. The full BF16 state takes $2 L H_{kv} d_h \times 2$~B/token, or $144$, $96$, and $256$~KiB/token for Qwen3-8B, Qwen3-30B-A3B, and Qwen3-32B, respectively~\cite{qwen3tech2025}. At $250$~Mbps a 32K prompt takes $154.6$~s, $103.1$~s, or $274.9$~s to transfer before protocol overhead.

With PP, each of the $P$ stages stores the cache for its $L_s$ layers. Only the boundary activations cross the link. For Qwen3-8B, Figure~\ref{fig:characterization}a gives $16$~KiB/token for native BF16 and $2.13$~KiB/token for \hintfour. Across the three models, the compressed payload is $67.8\text{--}96.4\times$ smaller than a full KV transfer.

\subsection{The Cost of Chunked Prefill}
\label{sec:motiv_coupling}

Prefill and decode contend for GPU time when they share a stage. A fixed chunk budget of $c$ tokens splits an $N$-token prompt into $k = \lceil N/c \rceil$ chunks~\cite{agrawal2024sarathi,hwang2026revisitpp}. We profiled 42 stage configurations on RTX~3090, RTX~4090, and PCIe A800 GPUs. The stage time for $N$ tokens at context length $C$ follows an affine model.
\begin{equation}
T_s(L_s, N, C) = L_s \bigl(\beta_0 + (\alpha_0 + \gamma_0 C) N\bigr),
\label{eq:affine}
\end{equation}
Here $\beta_0$ accounts for per-layer weight loads and dispatch. The per-token cost comprises $\alpha_0$ for feed-forward computation and $\gamma_0 C$ for attention over the context. Fixed and per-token costs balance at the roofline knee $N_{\text{knee}}(C) \triangleq \beta_0 / (\alpha_0 + \gamma_0 C)$.

Each token chunk crosses all $L_s$ layers and incurs the fixed cost again. With layer slicing, each of the $m$ slices runs $L_s/m$ layers at full width $N$. This incurs the fixed cost once per layer across the batch. For $k=N/c$ equal chunks, the model gives the following costs and relative overhead.
\begin{align}
T_s^{\mathrm{token}}(N, k) &= k L_s \beta_0 + L_s (\alpha_0 + \gamma_0 C) N, \label{eq:tok_cost}\\
T_s^{\mathrm{layer}}(N, m) &= L_s \beta_0 + L_s (\alpha_0 + \gamma_0 C) N, \label{eq:lay_cost}\\
\Delta_{\mathrm{tax}}(N, k, C) &= \frac{T_s^{\mathrm{token}} - T_s^{\mathrm{layer}}}{T_s^{\mathrm{layer}}} = \frac{k - 1}{1 + N / N_{\mathrm{knee}}(C)}. \label{eq:tax}
\end{align}

Figure~\ref{fig:characterization}b tests Eq.~\ref{eq:tax} on 4096-token batches. The knee is 70 tokens for the dense Qwen3-8B and 86 for Qwen3-32B. Chunks of 128--256 tokens add $25.2\%\text{--}63.8\%$ compute time on these models and up to $126.1\%$ with coding-agent context. Qwen3-30B-A3B routes each token to eight of 128 experts. Chunks of 128 tokens or more touch nearly all experts and stream about 30 billion weights. We measure $\beta_0 = 33.19$~ms and $\alpha_0 = 0.034$~ms/token, with a knee at 966 tokens. Using 512-token chunks adds $133.6\%$ to stage time. At 128 tokens, stage time rises from $173.9$~ms to $1,202.6$~ms, an increase of $591.7\%$.

At $c=4096$, a single prefill step on a 32K or 64K request can occupy an A800 stage for $519\text{--}643$~ms. Tail ITL reaches $627.2\text{--}3,232.3$~ms in panels b and c of Figure~\ref{fig:pareto_dilemma}. On the wire, native PP sends two BF16 tensors over one NCCL TCP stream. The $64$~MiB transfer for Qwen3-8B takes $2.15$~s at $250$~Mbps and $5.37$~s at $100$~Mbps. GPUs spend $80.6\%\text{--}89.2\%$ of wall time idle in \texttt{SendRecv}. With dynamic chunks, our \revisit baseline reaches $17.8$~tok/s and meets 2 of 9 requests' latency targets on the MoE workload. In the four-stage WAN run, it reaches a p95 TTFT of $53.9$~s and meets 1 of 21.

\subsection{Activation Redundancy and Packet Loss}
\label{sec:motiv_wire}

Wide layer slices still produce $N^* \times H$ activations at each stage boundary. vLLM sends both the hidden state $h$ and the residual $r$. The next Qwen3 layer consumes their sum through fused $\mathrm{RMSNorm}(h+r)$. Sending the sum $s=h+r$ and setting the residual to zero at the receiver halves the payload without loss.

Direct INT4 quantization is sensitive to outliers. At layer~20 of Qwen3-8B, the summed activations exhibit spikes greater than $18\times$ and kurtosis above 50. Following QuaRot~\cite{ashkboos2024quarot}, we apply Hadamard rotations to groups of 64 channels before quantization. The payload shrinks by $7.53\times$ and perplexity changes by less than $0.03\%$. Eager PyTorch takes $0.40\text{--}4.75$~ms per step for rotation and quantization, offsetting the transfer savings above $500$~Mbps. Panel c of Figure~\ref{fig:characterization} shows why we fuse these operations.

A single TCP stream is also sensitive to packet loss. On a $250$~Mbps link with $30$~ms RTT, throughput falls from $189.4$~tok/s to $42.0$~tok/s at $0.3\%$ loss and to $34.8$~tok/s at $0.5\%$ loss. Congestion-window reductions slow each transfer, and loss recovery can block later data on the stream.

%-------------------------------------------------------------------------------
\section{\sys Overview}
\label{sec:overview}
%-------------------------------------------------------------------------------

\begin{figure*}[t]
\centering
\includegraphics[width=\textwidth]{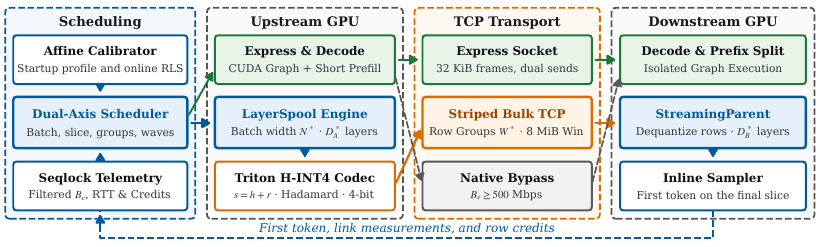}
\caption{\textbf{\sys on two pipeline stages.} The scheduler sends express work ahead of bulk work. The upstream GPU compresses completed activations, and the downstream GPU starts on arriving row groups. Token feedback and link measurements return to the scheduler.}
\label{fig:architecture}
\end{figure*}

The scheduler in Figure~\ref{fig:architecture} separates short prompts and decode from bulk prefill. It chooses a batch width $N^*$ to amortize weight loads over enough tokens. For each stage it chooses a slice depth $D_s^*$ between 1 and $L_s$ so decode can run between slices. It also chooses how many activation rows $W^*$ to send together. Startup calibration provides the first compute estimates. During serving the scheduler reads effective bandwidth, RTT, and transfer credits from shared memory.

On the upstream GPU, \layerspool advances bulk prefill by $D_A^*$ layers at a time and reuses its cached KV slot tables. The GPU runs decode on separate CUDA Graphs. After the final layer on that stage, the Triton kernel merges, rotates, and packs the activations into pinned host memory. The transport streams row groups over bulk TCP sockets, with places for express traffic to enter at every 32~KiB frame boundary. On fast, uncongested links the scheduler can send decode through native collectives.

At the receiving stage, \streamparent stages and dequantizes row groups as they arrive. \layerspool processes them at a depth $D_B^*$ chosen for that GPU. The final stage samples the first output token on the last slice and sends it back so the upstream scheduler can start decode.

%-------------------------------------------------------------------------------
\section{Prefill in Layer Slices}
\label{sec:layer}
%-------------------------------------------------------------------------------

\begin{figure*}[t]
\centering
\includegraphics[width=\textwidth]{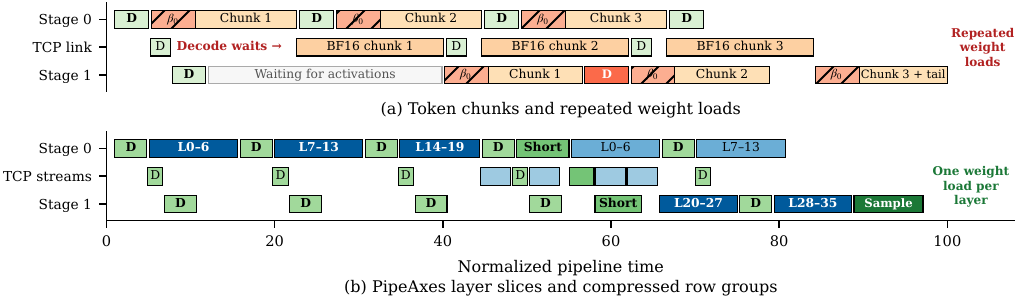}
\caption{\textbf{Execution across two pipeline stages.} In panel a, large BF16 sends delay the decode marked in red. Each chunk reloads weights on both GPUs. In panel b, dark blue marks the first prefill batch and lighter blue on stage 0 starts the next batch. The link carries light-blue row groups with room for decodes and short prefills between them. Blocks labeled D are decode steps. The final stage samples the first token when it finishes prefill.}
\label{fig:timeline}
\end{figure*}

\subsection{Resuming a Prefill Batch}
\label{sec:layer_spool}

In Figure~\ref{fig:timeline}b, a stage pauses an $N^*$-token batch after $D_s^*$ layers and resumes it after serving decodes. This avoids the repeated weight loads in Eq.~\ref{eq:tok_cost}. Slice budgets target preemption within $30\text{--}60$~ms. Rebuilding the batch and its attention metadata on every slice would add $2\text{--}6$~ms per slice.

Each GPU worker runs \layerspool as a FIFO with one active bulk batch. Before the first slice, \texttt{prepare\_layer\_prefill} stores the materialized input batch, RoPE positions, KV block tables, per-layer slot mappings, and the metadata for FlashAttention~\cite{dao2022flashattention} or FlashInfer~\cite{ye2025flashinfer} in a \texttt{LayerPrefill} descriptor. The active batch also retains two $N^* \times H$ BF16 tensors for the hidden state and residual. At $N^* = 4096$, they take $64$~MiB on 8B, $32$~MiB on 30B-A3B, and $80$~MiB on 32B, below $0.27\%$ of HBM. \layerspool uses the GPU time budget $B_{\text{gpu}}$ to choose the slice depth.
\begin{equation}
D_s^* = \max\!\left(1,\; \min\!\left(L_s - l_{\text{cur}},\; \left\lfloor \frac{B_{\mathrm{gpu}} \cdot L_s}{\hat{T}_s(L_s, N^*, C)} \right\rfloor\right)\right).
\label{eq:slice_depth}
\end{equation}
Under intra-node \texttt{TP2}, workers use a 4-byte CPU \texttt{MIN} all-reduce to agree on $D_s^*$. Non-blocking CUDA events record slice times. The worker runs slices back to back when no express work is pending. On intra-node PCIe stage boundaries without peer-to-peer access, \texttt{LocalBulkLink} uses \texttt{cudaHostRegister} to map \texttt{/dev/shm} into the CUDA host address space. A separate copy stream moves completed bulk batches while NCCL handles decode collectives.

\subsection{KV Dependencies and Decode Graphs}
\label{sec:layer_kv}

Decodes must wait for earlier prefill slices to populate the KV cache. Each bulk submission receives a monotonically increasing row watermark $W_{\text{sub}}(r)$. The scheduler admits request $r$ to decode once the downstream stage reports $W_{\text{cons}} \ge W_{\text{sub}}(r)$. A non-bulk step that touches a finished, cancelled, or overlapping request first calls \texttt{flush\_dependencies} to finish the active batch through layer $s_{\text{end}}$.

Variable prefill shapes require additional captured graphs or eager execution. Uniform decode batches can reuse graphs captured at supported batch sizes~\cite{chen2026muxwise}. \layerspool runs bulk prefill eagerly and replays full-stage CUDA Graphs for decode waves. When the express lane batches single-token decodes with a short non-terminal prefill, \texttt{split\_decode\_prefix} runs the decodes as a graph-compatible microbatch and follows it with the prefill.

\subsection{Terminal Inline Sampling}
\label{sec:layer_inline}

Some scheduling paths hold back the last 1--64 prompt tokens and sample the first output token on the express lane. The extra step and pipeline round trip add $30\text{--}150$~ms under load. \sys sets \texttt{tail\_rows = 0} and marks bulk batches that contain the end of a prompt in \texttt{awaiting\_first}. After the final stage finishes layer $L-1$, \texttt{finish\_layer\_prefill} runs the LM head and sampler on the terminal positions of $h+r$. This preserves prompt and output log probabilities. The sampled token goes directly to the upstream decode pool over the asynchronous ZeroMQ \texttt{ElasticFeedback} stream.

%-------------------------------------------------------------------------------
\section{Activation Compression and Transport}
\label{sec:wire}
%-------------------------------------------------------------------------------

\subsection{Fused Hadamard-INT4 Codec}
\label{sec:wire_codec}

The four-warp Triton~\cite{tillet2019triton} kernel \texttt{pp\_link\_quant} processes eight groups of $G=64$ channels per program instance. It loads $h, r \in \text{BF16}^{N \times H}$ and computes $s = \text{BF16}(\text{FP32}(h) + \text{FP32}(r))$ in registers. Setting the residual to zero at the receiver lets RMSNorm operate on the BF16-rounded sum.

The kernel rotates each group with the orthonormal Walsh--Hadamard transform $H_G$. This preserves $\|\tilde{s}_g\|_2 = \|s_g\|_2$ and spreads a spike of magnitude $|M|$ across the output channels, each with magnitude $|M|/\sqrt{G}$. Six butterfly stages gather channels at $\texttt{col} \oplus 2^j$ and normalize by $G^{-1/2}$. The kernel computes the per-group scale $\sigma_g = (\max_i |\tilde{s}_{g,i}| + 10^{-8})/7$, rounds and clamps the scaled values to $[-7,7]$, and packs two 4-bit values per byte. The payload starts with the FP16 scales.

Each row uses $H/2 + 2(H/64) = \frac{17}{32}H$~bytes. The two native BF16 tensors use 32 bits per channel, while the packed representation uses $4.25$, a $7.53\times$ reduction. The 8B, 30B-A3B, and 32B models use $2,176$, $1,088$, and $2,720$~B/token, respectively. The decoder unpacks the values and applies the inverse rotation $H_G^\top = H_G$. It writes BF16 in chunks of at most $2^{15}$ groups to limit scratch memory at $90\%$ KV occupancy. In Figure~\ref{fig:characterization}c, encoding and decoding take $0.100$~ms for $N \le 256$, a $3.98\times$ speedup over eager PyTorch. At $N=4096$ they take $0.948$~ms.

\subsection{Row-Group Streaming over TCP}
\label{sec:wire_stream}

At $250$~Mbps, sending an $8.5$~MiB compressed batch of 4096 tokens takes $285$~ms, exceeding the $100$~ms TPOT target. The ZeroMQ transport \pplink splits bulk transfers into groups of $W^*$ rows, with $64 \le W^* \le 1024$. Each group carries a JSON header in place of a blocking Gloo metadata exchange. \streamparent copies arriving groups to the GPU while earlier slices run.

Express and bulk traffic use separate sockets and $32$~KiB frames. One frame takes about 1~ms at $250$~Mbps. User space cannot interrupt data already in a kernel TCP buffer, so \pplink caps unacknowledged bulk data at $8$~MiB and can pace bulk sends by the ACK rate at $100$~Mbps and below. Express frames can pass bulk frames that are still queued in user space. Bulk groups rotate through parallel sockets so loss recovery on one socket does not block the others. When the link shows loss or jitter, \pplink sends small express frames on two sockets and filters duplicates with a 4,096-entry ring.

A full vLLM V2 GPU queue can block a host worker in CUDA while it holds Python's global interpreter lock. This can stall a sender that needs the same lock to finish an earlier transfer. We reproduced the deadlock with 48 prefills of 8K tokens each. \sys stages device-to-host copies in pinned memory and lets the sender wait on a CUDA event. The worker waits only for ZeroMQ to acknowledge that it queued the handoff.

\subsection{Bandwidth Estimation and Native Bypass}
\label{sec:wire_telemetry}

Preempted and paced bulk transfers report less throughput than the link can carry. Using those samples can make the scheduler lower its bandwidth estimate even when the physical link rate stays constant. \pplink measures the rolling p20 RTT for frames of at most $8$~KiB. It updates effective bandwidth $B_e$ over $200$~ms windows using only uncontended and unpaced payloads with a native equivalent of at least $256$~KiB. Its exponential moving average weights downward changes by $0.50$ and upward changes by $0.20$. The scheduler reads $B_e$, RTT, $W_{\text{cons}}$, and $\gamma_s$ through an eight-word seqlock in \texttt{/dev/shm}. These reads take less than $1\,\mu$s and require no IPC system calls.

At $250$~Mbps, \pplink selects \hintfour in $99.95\%$ of steps. Above $500$~Mbps it can use INT8 or BF16 for small uncontended transfers. For a pure decode wave with $N=M_{\text{req}}$, the scheduler can use native collectives when $B_e \ge 500$~Mbps and $W_{\text{sub}} = W_{\text{cons}}$. It takes the native path if its predicted latency is at most that of \pplink.
\begin{equation}
T_{\mathrm{native}}(N) = 0.35 + 0.5\,\mathrm{RTT} + \frac{4 H N}{125 B_e} \;\le\; T_{\mathrm{link}}(N).
\label{eq:bypass}
\end{equation}

%-------------------------------------------------------------------------------
\section{Scheduling and Admission Control}
\label{sec:sched}
%-------------------------------------------------------------------------------

\subsection{Scheduling with Latency Slack}
\label{sec:sched_alg}

\begin{figure}[t]
\centering
\scriptsize
\renewcommand{\arraystretch}{1.05}
\setlength{\tabcolsep}{2pt}
\begin{tabular}{@{}p{0.99\columnwidth}@{}}
\toprule
\textbf{Algorithm 1.} Scheduling at step $t$ \\
\midrule
\textbf{Input:} Queues $(Q_{\text{run}}, Q_{\text{wait}})$, stage profiles $\{StageCost_s\}_{s=0}^{P-1}$, seqlock telemetry $(B_e, \mathrm{RTT}, W_{\text{cons}}, \gamma_s)$, target $\mathrm{ITL}_{\text{SLO}}$. \\
1: Refresh $(B_e, \mathrm{RTT}, W_{\text{cons}})$; retire watermarks $\le W_{\text{cons}}$. \\
2: Partition runnable requests into decode $Q_{\text{dec}}$, express $Q_{\text{exp}}$, and bulk $Q_{\text{bulk}}$. Express uses prompts of at most $N_{\text{exp}}$ tokens with TTFT targets of at most 2~s. Reserve at least $25\%$ of \texttt{max\_seqs} slots for express. \\
3: \textbf{if} $Q_{\text{bulk}} = \emptyset \land W_{\text{sub}} = W_{\text{cons}} \land B_e \ge 500$ \textbf{then} \\
4: \quad Schedule hybrid wave + check Eq.~\ref{eq:bypass} Native Bypass; \textbf{return}. \\
5: Set $T_{\text{cyc}} \leftarrow \mathrm{RTT} + \sum_{s=0}^{P-1} T_s^{\text{dec}}(|Q_{\text{dec}}|)$, oldest age $a_{\max}$, and slack $S \leftarrow \mathrm{ITL}_{\text{SLO}} - a_{\max} - T_{\text{cyc}}$. \\
6: Set row wire cost $\tau_w \leftarrow \frac{9 H / 16}{125 B_e}$; select wire group $W^* \leftarrow \mathrm{align}_{64}\!\bigl(\min(1024, \lfloor \min(16, \mathrm{ITL}_{\text{SLO}}/12) / \tau_w \rfloor)\bigr)$. \\
7: Set $N_{\text{fill}} \leftarrow \lfloor \max(0.25\,\mathrm{ITL}_{\text{SLO}}, 0.60\,\mathrm{ITL}_{\text{SLO}} - 0.5\,\mathrm{RTT}) / \tau_w \rfloor$ and $N^* \leftarrow \mathrm{align}_{64}\!\bigl(\min(N_{\max}, \max(256, N_{\text{fill}}, \min(512, \max_s N_{\text{knee}, s})))\bigr)$. \\
8: Set credit cap $C_{\text{cred}} \leftarrow \max(P \cdot N^*, N^* + \lfloor \mathrm{RTT} / \tau_w \rfloor)$. \\
9: \textbf{if} $Q_{\text{bulk}} \neq \emptyset \land (W_{\text{sub}} - W_{\text{cons}} + N^* \le C_{\text{cred}}) \land \mathrm{BulkDue}(t, a_{\max})$ \textbf{then} \\
10: \quad Admit an $N^*$-row batch with \texttt{tail\_rows=0}. Use $B_{\text{gpu}} \leftarrow \min(0.35\,\mathrm{ITL}_{\text{SLO}}, 0.5\,\min(S, 0.5\,\mathrm{ITL}_{\text{SLO}} - T_{\text{cyc}}))$ in Eq.~\ref{eq:slice_depth}. \\
11: \textbf{else if} $Q_{\text{dec}} \neq \emptyset$ \textbf{then} schedule decode, balancing waves if $P > 2$. \\
12: \textbf{else} Admit $Q_{\text{exp}}$ under the policy in Section~\ref{sec:sched_pp4}. \\
\bottomrule
\end{tabular}
\end{figure}

The scheduler reserves $\max(1,\lfloor\text{max\_seqs}/4\rfloor)$ slots for express requests so 32K--64K prefills cannot occupy all slots. Prompts shorter than 512 tokens can use this reservation. On local PCIe, $N_{\max}=4096$. Over WAN, the scheduler chooses $N^*$ using the largest stage knee and the time needed to fill the link. The credit cap $C_{\text{cred}}$ lets up to the larger of $P$ batches or one bandwidth-delay product plus one batch remain in flight.

The wire-group size $W^*$ limits estimated serialization time to $\min(16\text{ ms},\mathrm{ITL}_{\text{SLO}}/12)$. The estimate budgets $9H/16$~bytes per row, including headroom above the packed \hintfour payload of $17H/32$~bytes. For the GPU slice, the scheduler allocates as much as half the remaining decode slack, subject to the time budget left in half an ITL period and a cap of $0.35\,\mathrm{ITL}_{\text{SLO}}$. Each stage converts the budget $B_{\text{gpu}}$ to a layer count with Eq.~\ref{eq:slice_depth}.

Consider an 8K prefill and 16 decodes on Qwen3-8B with a 200~ms ITL target. An RTX~4090 runs the first 20 layers and an RTX~3090 runs the last 16. At $250$~Mbps with $30$~ms RTT, the scheduler estimates $\tau_w=0.0737$~ms per row and $T_{\text{cyc}}\approx48$~ms. It budgets $14.2$~ms for a 192-row group. The packed payload takes $13.4$~ms at line rate. For a 1024-row batch and a $26$~ms slice budget, the 4090 takes $54$~ms for a full stage and runs nine layers per slice. The 3090 takes $98$~ms for a full stage and runs four layers per slice. Decode traffic can use the link within $1$~ms of a frame boundary, and the decode graph takes $9$~ms.

\subsection{Latency Model Calibration}
\label{sec:sched_cal}

At startup, \sys calibrates its latency model in less than 2~s. It measures prefill at $N_1=256$ and $N_2=2048$ with no prior context, and decode at batch sizes of 8 and 32. The prefill measurements give $\alpha_0=[T(N_2)-T(N_1)]/(N_2-N_1)$ and $\beta_0=T(N_1)-\alpha_0 N_1$.

Whenever \layerspool completes a batch with prior context $C>0$, the worker updates the attention slope $\gamma_0$ by scalar RLS~\cite{haykin2014adaptive}. It uses feature $f=N C/32768$ and forgetting factor $\lambda=0.98$, then publishes $\gamma_0$ through \texttt{LinkTelemetry}. Section~\ref{sec:eval_ablate} measures the effect of calibration errors on SLO attainment.

\subsection{Decode Waves and Deadline-Based Admission}
\label{sec:sched_pp4}

When all active decodes enter a single microbatch, only one stage works on them at a time. Once $|Q_{\text{dec}}| \ge 2P$, \sys distributes the decodes evenly across $W_{\text{waves}} = \max(1,P-\mathbb{I}[\text{has\_bulk}])$ waves. This reserves a wave slot for \layerspool when bulk prefill is pending.

Under heavy four-stage load, a short request at the head of the queue can consume an admission slot after its TTFT deadline is already out of reach. \sys estimates the remaining slack as $t_{\text{arr}}+2.0-t_{\text{now}}-\hat{t}_{\text{exec}}$. When this quantity is negative for the head, \texttt{select\_express\_waiting} can admit a short request that still has time to meet its deadline. It permits at most two consecutive bypasses before serving the oldest request. Section~\ref{sec:eval_topo} measures the effect on goodput.

\subsection{Implementation}
\label{sec:impl}

Our implementation uses vLLM v0.23.0 at commit \texttt{0fc695fc}. It modifies 2,334 lines in the engine and workers and adds 1,789 lines of standalone Python and Triton code.

%-------------------------------------------------------------------------------
\section{Evaluation}
\label{sec:eval}
%-------------------------------------------------------------------------------

\subsection{Experimental Setup}
\label{sec:eval_setup}

Our datacenter testbed uses two NVIDIA A800s with 80~GB each on PCIe Gen4 without NVLink or peer-to-peer access. It runs Ubuntu 22.04, CUDA 12.8, and PyTorch 2.11. Our cross-node testbed connects a host with two 24~GB RTX~4090s to a host with two 24~GB RTX~3090s over 1~GbE. A layer takes about $2.1\times$ as long on the 3090. We use Linux \texttt{tc netem} to set link rates from 100 to 1000~Mbps, with 30~ms RTT, 5~ms jitter, and $0.3\%\text{--}0.5\%$ loss in the corresponding tests.

All models use BF16 weights. We split Qwen3-8B into stages of 20 and 16 layers for PP2 or 10, 10, 8, and 8 layers for PP4. For Qwen3-30B-A3B, we place 24 layers on each stage on A800 and in the hybrid PP2$\times$TP2 setup, and 12 layers on each stage for PP4. Qwen3-32B uses two stages of 32 layers on A800 and serves up to 64K tokens with YaRN~\cite{peng2024yarn}.

\begin{table*}[t]
\centering
\footnotesize
\renewcommand{\arraystretch}{0.93}
\setlength{\tabcolsep}{3.6pt}
\caption{\textbf{End-to-end results on seven datacenter and WAN workloads.} Systems within each workload replay the same requests. A request meets its SLO when it meets all three latency targets. TTFT covers short requests where present and all requests otherwise. Tail ITL is the p90 across requests of each request's p95 ITL. E2E measures end-to-end request latency.}
\label{tab:main_all_metrics}
\begin{tabular}{@{}llrrrrrrr@{}}
\toprule
\textbf{Scenario \& Topology} & \textbf{System} & \textbf{SLO Met} & \textbf{Goodput} & \textbf{Output} & \textbf{TTFT$_{p95}$} & \textbf{TPOT$_{p95}$} & \textbf{ITL$_{p95,P90}$} & \textbf{E2E$_{p95}$} \\
 & & \textbf{Met/Total} & \textbf{req/s $\uparrow$} & \textbf{tok/s $\uparrow$} & \textbf{ms $\downarrow$} & \textbf{ms $\downarrow$} & \textbf{ms $\downarrow$} & \textbf{s $\downarrow$} \\
\midrule
\multirow{4}{*}{Local PP2 · $2\times$4090 · 8B · 6.45 req/s}
 & \textbf{\sys} & \textbf{135/188} & \textbf{2.65} & 472.3 & \textbf{2,323} & \textbf{63.9} & 108.2 & \textbf{8.51} \\
 & Fixed-512 + Express & 112/188 & 2.16 & 464.3 & 8,234 & 82.7 & 108.4 & 19.11 \\
 & \revisit & 82/188 & 1.27 & 373.8 & 13,043 & 89.6 & \textbf{87.5} & 22.40 \\
 & Native vLLM V2 & 14/188 & 0.29 & \textbf{495.6} & 15,016 & 67.5 & 463.5 & 21.14 \\
\midrule
\multirow{3}{*}{Local PP2 · $2\times$A800 · 30B-A3B · 32K}
 & \textbf{\sys} & \textbf{9/9} & \textbf{0.30} & 38.1 & 6,706 & \textbf{74.5} & \textbf{111.1} & \textbf{14.09} \\
 & Fixed-4096 + Express & 4/9 & 0.14 & \textbf{39.3} & \textbf{5,478} & 99.6 & 627.2 & 14.94 \\
 & \revisit & 2/9 & 0.03 & 17.8 & 41,871 & 109.5 & 289.0 & 43.18 \\
\midrule
\multirow{3}{*}{Local PP2 · $2\times$A800 · Dense 32B · 64K YaRN}
 & \textbf{\sys} & \textbf{15/23} & \textbf{0.074} & 14.6 & \textbf{512} & \textbf{77.8} & \textbf{128.4} & \textbf{10.14} \\
 & \revisit & 1/23 & 0.003 & 10.1 & 264,134 & 2,129.2 & 2,707.4 & 279.80 \\
 & Fixed-4096 + Express & 0/23 & 0.000 & \textbf{15.6} & 158,163 & 1,290.5 & 3,232.3 & 174.61 \\
\midrule
\multirow{3}{*}{WAN 4090$\rightarrow$3090 · 8B · 250M · 2.62 req/s}
 & \textbf{\sys} & \textbf{75/77} & \textbf{1.60} & \textbf{210.3} & \textbf{1,156} & \textbf{93.2} & 172.8 & \textbf{12.90} \\
 & Best Static, Fixed-256 & 36/77 & 0.76 & 207.7 & 2,181 & 131.0 & 137.5 & 16.84 \\
 & \revisit & 42/77 & 0.74 & 173.9 & 1,984 & 132.3 & \textbf{95.1} & 17.91 \\
\midrule
\multirow{3}{*}{WAN 4090$\rightarrow$3090 · 8B · 10$\times$ BW Swing}
 & \textbf{\sys} & \textbf{74/74} & \textbf{1.75} & \textbf{223.9} & \textbf{802} & \textbf{93.5} & 164.1 & \textbf{12.15} \\
 & \revisit & 46/74 & 0.94 & 193.5 & 1,840 & 116.7 & \textbf{96.7} & 15.39 \\
 & BW-Oracle Static & 3/74 & 0.07 & 223.4 & 3,847 & 124.0 & 301.2 & 18.92 \\
\midrule
\multirow{3}{*}{WAN 3090$\rightarrow$4090 Reverse · 8B · 250M}
 & \textbf{\sys} & \textbf{70/77} & \textbf{1.37} & 192.4 & \textbf{780} & \textbf{86.4} & 177.1 & \textbf{11.77} \\
 & \revisit & 40/77 & 0.67 & 165.1 & 2,004 & 148.8 & \textbf{129.3} & 20.02 \\
 & Fixed-512 + Express & 25/77 & 0.54 & \textbf{211.9} & 4,107 & 131.0 & 222.1 & 18.80 \\
\midrule
\multirow{3}{*}{WAN PP4 · MoE 30B-A3B · 32K · 250M}
 & \textbf{\sys} & \textbf{16/21} & \textbf{0.28} & \textbf{47.1} & \textbf{509} & \textbf{85.7} & \textbf{130.3} & \textbf{11.08} \\
 & Fixed-4096 + Express & 4/21 & 0.06 & 42.3 & 8,260 & 282.9 & 767.2 & 42.08 \\
 & \revisit & 1/21 & 0.01 & 31.7 & 53,932 & 460.6 & 1,017.7 & 63.27 \\
\bottomrule
\end{tabular}
\end{table*}

All baselines use the same vLLM v0.23.0 V2 engine and kernel stack. We restart the engine between runs. Native vLLM V2 uses 4096-token chunks, CUDA Graphs, and NCCL over TCP. For \revisit, we port the predictive chunking and delay scheduling of Hwang and Ahn~\cite{hwang2026revisitpp} from SGLang~\cite{zheng2024sglang} to \texttt{revisit\_pp.py} in vLLM. The port calibrates each GPU and chooses among 128, 256, 512, 1024, 2048, and 4096 tokens up to the configured maximum. On WAN it uses the same \hintfour priority transport as \sys.

The static grid uses chunk sizes of 256, 512, 1024, and 4096 tokens with \hintfour and a separate express socket. We compare against the best static run at each arrival rate across the 16 runs in the WAN sweep. On the bandwidth swing, BW-Oracle Static knows the changes in advance.

Across 97 paired replays covering 5,016 requests, goodput counts only requests that meet all three SLOs. TTFT must stay within 2~s for short requests and 12~s for long requests. Both limits double for the multi-turn coding-agent workloads. TPOT must stay within 100~ms and each request's p95 ITL within 200~ms.

\subsection{Single-Node PCIe Serving}
\label{sec:eval_dc}

\begin{figure}[t]
\centering
\includegraphics[width=\columnwidth]{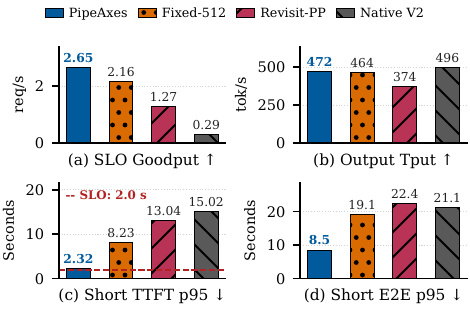}
\caption{\textbf{Qwen3-8B on two PCIe RTX~4090s.} The local replay submits 188 requests at 6.45~req/s. Panels c and d report p95 latency for short requests.}
\label{fig:local_dc}
\end{figure}

\begin{figure}[t]
\centering
\includegraphics[width=\columnwidth]{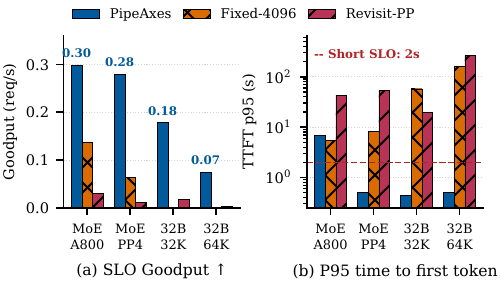}
\caption{\textbf{Goodput and first-token latency on long prompts.} The PCIe workloads run on two A800s, and the WAN workload uses four pipeline stages. Panel b plots p95 TTFT for short requests where present and for all requests otherwise. The axis is logarithmic.}
\label{fig:long_context}
\end{figure}

At $6.45$~req/s on two RTX~4090s, native vLLM V2 produces $495.6$~tok/s, but its tail ITL reaches $463.5$~ms and only 14 of 188 requests meet their SLOs. \revisit reduces tail ITL to $87.5$~ms with smaller chunks. Repeated weight loads bring output down to $373.8$~tok/s, and short-request p95 TTFT reaches $13.04$~s. In Figure~\ref{fig:local_dc}, \sys retains $95.3\%$ of native output at $472.3$~tok/s, with $108.2$~ms tail ITL and $2.32$~s p95 TTFT for short requests. Its goodput of $2.65$~req/s is $9.19\times$ native vLLM, $2.08\times$ \revisit, and $1.23\times$ Fixed-512. At $4.79$~req/s, all 142 requests meet their SLOs.

On the 32K MoE workload, Fixed-4096 reaches a tail ITL of $627.2$~ms and meets the targets for four of nine requests. \revisit meets the targets for two and runs at $17.8$~tok/s, while Row-Slice Only meets them for one. Figure~\ref{fig:long_context} shows \sys meeting all nine with $111.1$~ms tail ITL. Its goodput of $0.30$~req/s is $9.61\times$ \revisit and $2.18\times$ Fixed-4096. It preserves $96.9\%$ of Fixed-4096's output at $38.1$~tok/s.

On dense Qwen3-32B with mixed 32K contexts, \sys meets the SLOs for all 24 short requests and reaches $9.89\times$ the goodput of \revisit. Its p95 TTFT for short requests is $439$~ms, compared with $19.6\text{--}57.3$~s across the baselines.

The 64K YaRN workload in Table~\ref{tab:main_all_metrics} includes 15 short requests and eight prompts of 56K--63K tokens. Fixed-4096, Fixed-512, and native vLLM V2 miss all 23 SLOs with $129.8\text{--}183.4$~s p95 TTFT for short requests. \revisit meets one SLO and reaches $264.1$~s p95 TTFT. \sys meets all 15 short requests' targets. It brings short-request p95 TTFT down to $512$~ms and tail ITL to $128.4$~ms while retaining $93.6\%$ of Fixed-4096's output at $14.6$~tok/s.

\subsection{Serving Capacity over a WAN}
\label{sec:eval_wan}

\begin{figure*}[t]
\centering
\includegraphics[width=\textwidth]{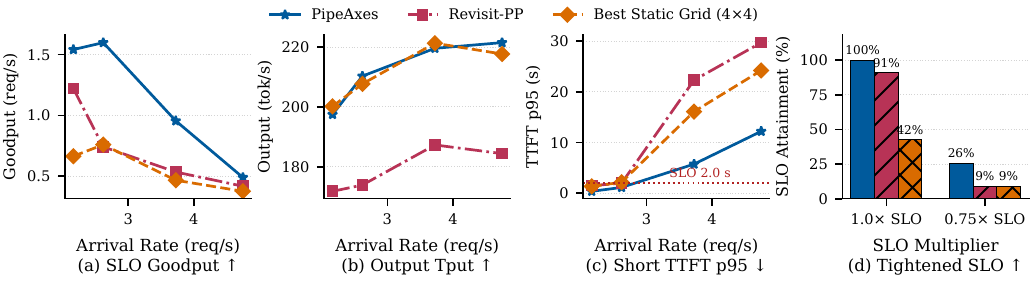}
\caption{\textbf{Qwen3-8B over a 250~Mbps WAN link.} Traffic flows from an RTX~4090 to an RTX~3090 at 30~ms RTT. Panels a through c compare \sys with calibrated \revisit and the best fixed chunk size at each arrival rate. Panel d tightens all latency targets by $25\%$ at 2.17~req/s.}
\label{fig:capacity}
\end{figure*}

We vary the offered load from $2.17$ to $4.75$~req/s in Figure~\ref{fig:capacity}. \sys delivers $1.30\text{--}2.33\times$ the goodput of the best static run at each rate. At $2.62$~req/s, \sys reaches $1.60$~req/s and the best static run reaches $0.76$~req/s. Compared with calibrated \revisit, \sys serves $1.17\text{--}2.16\times$ as many requests within their latency targets and produces $15\%\text{--}21\%$ more tokens per second. At $3.73$~req/s, it reduces short-request p95 TTFT from $22.34$~s to $5.76$~s.

At $2.17$~req/s, tightening the targets to $1.5$~s TTFT for short requests, $75$~ms TPOT, and $150$~ms ITL changes the best static chunk size from 256 to 1024 tokens. In panel d of Figure~\ref{fig:capacity}, both baselines meet 6 of 66 requests' SLOs, or $9.1\%$. \sys meets $25.8\%$, about $2.83\times$ as many.

We also compare against unmodified vLLM in 12 uncached and 40 cached configurations over $100\text{--}1000$~Mbps. The geometric mean speedup is $1.97\times$ and reaches $6.52\times$ on 48 prefills of 8K tokens each. At $250$~Mbps, the two stages spend $40.8\%$ and $38.6\%$ of their time waiting, down from $89.2\%$ and $80.6\%$.

\subsection{Multi-Turn Coding Agents}
\label{sec:eval_agent}

\begin{figure*}[t]
\centering
\includegraphics[width=\textwidth]{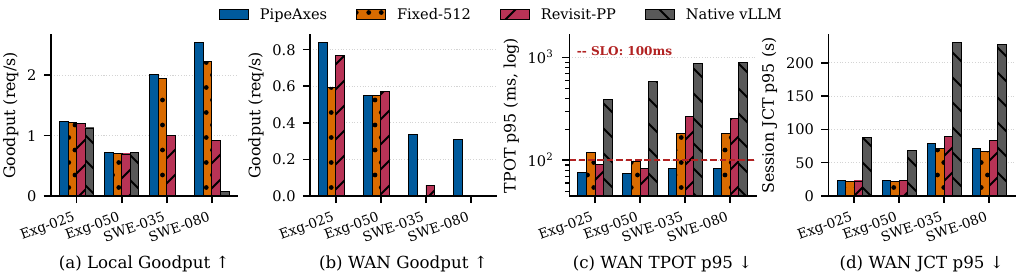}
\caption{\textbf{Coding-agent replays.} The Exgentic trace has 60 calls. The SWE-agent trace has 90 calls across 18 sessions. Panel a uses two local RTX~4090s. Panels b through d run from a 4090 to a 3090 over 250~Mbps at 30~ms RTT. Panel c shows p95 TPOT on a log scale.}
\label{fig:agents}
\end{figure*}

We extract request lengths and timing from Exgentic~\cite{exgentic2026traces}. For SWE-agent~\cite{yang2024sweagent}, we take request lengths from the trajectories and space turns 2~s apart. On the local testbed in Figure~\ref{fig:agents}a, native vLLM V2 meets zero and three of 90 requests' SLOs across the two SWE-agent workloads. \sys meets the targets for all 60 Exgentic calls and all 90 calls in SWE-080. On SWE-080, it achieves $2.54$~req/s goodput, or $2.78\times$ that of \revisit, and reduces p95 session completion time from $37.4$~s to $31.2$~s.

Over WAN, Fixed-512 completes SWE-agent sessions in $66.6\text{--}71.4$~s at p95, compared with $70.9\text{--}78.7$~s for \sys. Under Fixed-512, p95 TPOT reaches $180.7\text{--}181.1$~ms and no SWE-agent request meets all latency targets. With \revisit, p95 TPOT reaches $256\text{--}266$~ms and SLO attainment is 5 of 90 on SWE-035 and zero on SWE-080. \sys keeps p95 TPOT at $83.5\text{--}84.0$~ms. It delivers $6.13\times$ \revisit goodput on SWE-035 and cuts p95 session completion time by $10.7\text{--}12.1$~s across the two SWE-agent workloads.

\subsection{Pipeline Topologies}
\label{sec:eval_topo}

\begin{figure}[t]
\centering
\includegraphics[width=\columnwidth]{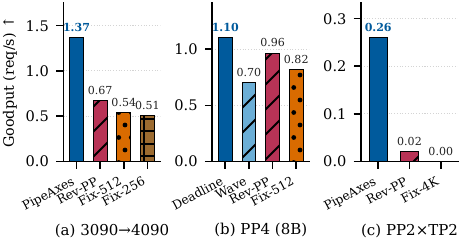}
\caption{\textbf{Pipeline topologies at 250~Mbps and 30~ms RTT.} Panel a sends Qwen3-8B activations from a 3090 to a 4090. Panel b serves 142 requests on four Qwen3-8B stages and compares admission policies for \sys. Panel c uses two pipeline stages with two-way tensor parallelism for the 32K MoE workload.}
\label{fig:topologies}
\end{figure}

With the RTX~3090 upstream in Figure~\ref{fig:topologies}a, \sys meets 70 of 77 requests' latency targets and reaches $1.37$~req/s goodput. This is $2.04\times$ \revisit and $2.54\times$ Fixed-512. Short-request p95 TTFT falls from $2.00$~s with \revisit to $780$~ms. At an offered load of $3.56$~req/s, \sys reaches $7.43\times$ the goodput of \revisit.

Across four stages on the 32K MoE workload, \sys meets 16 of 21 requests' targets. Its $0.28$~req/s goodput is $4.44\times$ Fixed-4096 and $23.3\times$ \revisit. Table~\ref{tab:main_all_metrics} reports p95 TTFT of $509$~ms for \sys and $53.9$~s for \revisit. On the overloaded four-stage 8B workload, deadline-based admission increases goodput from $0.70$ to $1.10$~req/s over wave balancing alone. \revisit reaches $0.96$~req/s and Fixed-512 reaches $0.82$~req/s.

With two pipeline stages and two-way tensor parallelism, \sys synchronizes the two transfer links. On the 32K MoE workload in Figure~\ref{fig:topologies}c, it reaches $0.260$~req/s, meets 16 of 21 requests' targets, and brings p95 TTFT to $389$~ms. \revisit achieves $0.021$~req/s and Fixed-4096 meets no SLOs.

\subsection{Bandwidth Changes and Packet Loss}
\label{sec:eval_dyn}

\begin{figure}[t]
\centering
\includegraphics[width=\columnwidth]{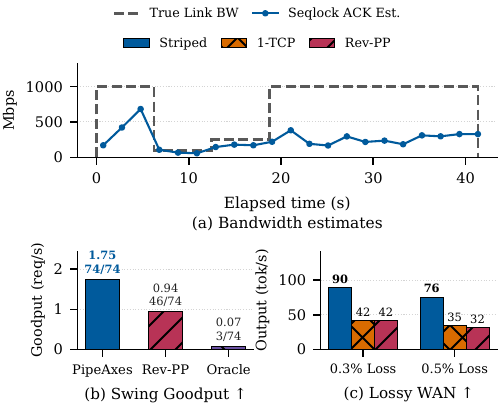}
\caption{\textbf{Bandwidth changes and packet loss on Qwen3-8B at 30~ms RTT.} Panels a and b show estimated bandwidth and goodput as the link steps through 1000, 100, 250, and 1000~Mbps. Panel c measures output at 250~Mbps with $0.3\%$ and $0.5\%$ packet loss.}
\label{fig:dyn_lossy}
\end{figure}

\begin{figure}[t]
\centering
\includegraphics[width=\columnwidth]{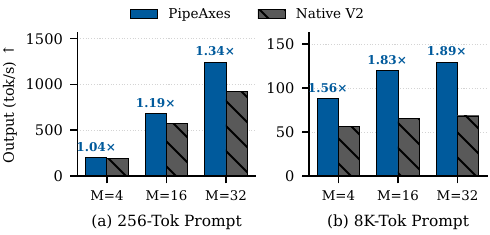}
\caption{\textbf{Native bypass on two RTX~4090s over 1~GbE.} The Qwen3-8B runs compare output throughput at concurrency levels of 4, 16, and 32.}
\label{fig:fastpath}
\end{figure}

In Figure~\ref{fig:dyn_lossy}, we change the link rate about every six seconds. The estimator follows the drop from 1000 to 100~Mbps without oscillating. It stays below line rate during the 1~Gbps phases. The oracle baseline enters the 100~Mbps phase with transfers from its prior 1024-token chunks still queued. It meets 3 of 74 requests' targets and achieves $0.07$~req/s. \sys meets all 74 targets and achieves $1.75$~req/s, or $1.86\times$ \revisit goodput and $24.6\times$ the oracle's.

TCP striping with duplicate express sends sustains $89.9$~tok/s at $0.3\%$ loss and $75.9$~tok/s at $0.5\%$ loss. Across the two loss rates, this is $2.14\text{--}2.18\times$ single-stream TCP and $2.17\text{--}2.36\times$ \revisit output. At $0.3\%$ loss, tail ITL falls from $4,087$~ms to $399.5$~ms.

On the 1~GbE link in Figure~\ref{fig:fastpath}, native bypass skips staging when Eq.~\ref{eq:bypass} favors direct transfer. With 32 concurrent requests, \sys reaches $1,242.0$~tok/s on 256-token prompts and $129.5$~tok/s on 8K prompts. These rates are $1.34\times$ and $1.89\times$ native vLLM V2. With 16 concurrent 8K prompts, \sys meets the SLOs for 13 requests and native vLLM V2 meets none.

\subsection{Ablation, Model Quality, and Overhead}
\label{sec:eval_ablate}

\begin{figure}[t]
\centering
\includegraphics[width=\columnwidth]{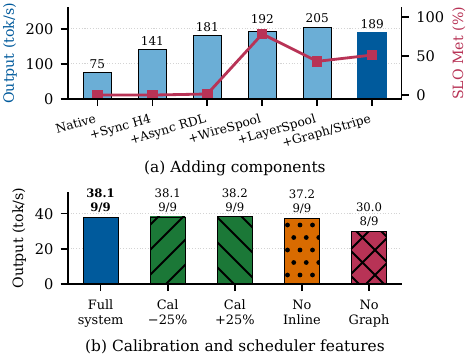}
\caption{\textbf{Component ablations and calibration sensitivity.} Panel a adds components in sequence on two RTX~3090s with 74 requests at 250~Mbps. Panel b varies calibration by up to $25\%$ in either direction and disables scheduler features on the 32K MoE workload on two A800s.}
\label{fig:ablation}
\end{figure}

\begin{table}[t]
\centering
\scriptsize
\setlength{\tabcolsep}{3.5pt}
\caption{\textbf{Model quality on \texttt{Qwen3-8B}.} We use the WikiText-2~\cite{merity2017wikitext}, MMLU~\cite{hendrycks2021mmlu}, and GSM8K~\cite{cobbe2021gsm8k} test sets, the four-option ARC-Challenge~\cite{clark2018arc} test questions, and 24 key-value retrieval prompts derived from SCBench~\cite{li2025scbench} at 32K tokens.}
\label{tab:quality}
\begin{tabular}{@{}lrrrr@{}}
\toprule
\textbf{Benchmark} & \textbf{BF16} & \textbf{Merge} & \textbf{INT8} & \textbf{\sys \hintfour} \\
\midrule
WikiText-2 PPL, 299K tok $\downarrow$ & 8.688 & 8.681 & 8.681 & \textbf{8.686} \\
MMLU, 14,042 items $\uparrow$ & 75.02\% & 75.00\% & 74.98\% & \textbf{75.08\%} \\
ARC-C, 1,165 items $\uparrow$ & 91.50\% & 91.59\% & 91.50\% & 91.42\% \\
GSM8K, 1,319 items $\uparrow$ & 90.30\% & 89.92\% & 89.69\% & 89.84\% \\
SCBench, 32K context $\uparrow$ & 15/24 & --- & --- & \textbf{16/24} \\
\bottomrule
\end{tabular}
\end{table}

\begin{table}[t]
\centering
\scriptsize
\setlength{\tabcolsep}{3.2pt}
\caption{\textbf{Runtime and memory costs.} Results come from 79 scheduler traces and microbenchmarks on RTX~4090 and A800. Scheduler times list p50 and p95 in that order.}
\label{tab:overhead}
\begin{tabular}{@{}lrr@{}}
\toprule
\textbf{Component} & \textbf{Cost} & \textbf{Comparison / Share} \\
\midrule
Scheduler, local 8B & $0.20 / 0.51$~ms & $0.86\times$ native at $0.23$~ms \\
Scheduler, A800 32B & $0.07 / 0.18$~ms & $0.57\times$ fixed at $0.13$~ms \\
Scheduler, WAN 8B & $0.36 / 1.25$~ms & $<0.4\%$ of decode cycle \\
Triton \hintfour for $N \le 256$ & $0.100$~ms & $3.98\times$ PyTorch INT4 speed \\
\texttt{/dev/shm} Seqlock read/write & $<1\,\mu$s & Lock-free 64-B mmap \\
\layerspool HBM at $N^*=4096$ & $32\text{--}80$~MiB & $0.10\%\text{--}0.27\%$ of HBM \\
\bottomrule
\end{tabular}
\end{table}

In Figure~\ref{fig:ablation}a, \hintfour with asynchronous transport raises output from $75.2$ to $180.6$~tok/s. Only one of 74 requests meets its SLO. Adding \streamparent lets 58 of 74 meet their targets and raises output to $192.5$~tok/s. \layerspool raises output to $205.2$~tok/s. With decode graphs isolated and TCP striped, tail ITL under overload is $194.3$~ms. On a separate 100~Mbps run from a 4090 to a 3090, disabling \layerspool reduces SLO attainment from 32 of 106 requests to zero.

In Figure~\ref{fig:ablation}b, shifting calibration by $25\%$ in either direction leaves all nine requests within their SLOs and output at $38.1\text{--}38.2$~tok/s on the 32K MoE workload. Disabling inline sampling raises p95 TPOT from $73.7$ to $86.6$~ms. Disabling decode graph isolation reduces goodput by $30.2\%$.

On MMLU, McNemar's test gives $p=1.0$ for residual merging. With \hintfour and \layerspool forced on, MMLU accuracy rises by 0.06 percentage points and $98.1\%$ of predictions agree with BF16. GSM8K accuracy falls by 0.46 percentage points with $94.99\%$ agreement. Table~\ref{tab:quality} also shows WikiText-2 perplexity of $8.6860$ with compression and $8.6882$ with BF16. On the 32K SCBench subset, \sys answers 16 of 24 correctly and BF16 answers 15. Table~\ref{tab:overhead} gives median scheduler times of $0.07\text{--}0.36$~ms and HBM overhead of at most $0.27\%$.

%-------------------------------------------------------------------------------
\section{Discussion and Limitations}
\label{sec:discuss}
%-------------------------------------------------------------------------------

At $3.73$~req/s on a 100~Mbps link, wide prefill batches can produce activations faster than the link can carry them. Pacing bulk traffic at the ACK rate brings p95 TTFT for short requests to $6.41$~s, down from $29.53$~s with \revisit. In a separate run, pairing batches narrower than 256 tokens raises goodput from $0.42$ to $0.80$~req/s. We have yet to combine the two policies below 100~Mbps.

On two GPUs, computing the prefill for a 60K prompt already takes over 14~s. That exceeds the 12~s TTFT target even though all 15 concurrent short requests meet their SLOs. Elastic sequence parallelism~\cite{wu2024loongserve} could distribute this prefill work across more GPUs. The prototype supports dense and MoE Qwen3 models. We disable prefix caching to account for all prefill computation. Adding prefix-hit handling and supporting other model families are extensions to the prototype.

Coding agents helped with experiment harnesses, plotting scripts, and routine refactoring. We directed and verified the designs, deadlock diagnoses, kernel formulations, and runs on the physical testbeds.

%-------------------------------------------------------------------------------
\section{Related Work}
\label{sec:related}
%-------------------------------------------------------------------------------

\begin{table}[t]
\centering
\scriptsize
\setlength{\tabcolsep}{2.6pt}
\caption{\textbf{Architectural comparison} of the designs described in each paper. Wire preemption and loss or bandwidth adaptation refer to inter-stage activation transfers.}
\label{tab:related}
\begin{tabular}{@{}lccccc@{}}
\toprule
\textbf{System} & \textbf{Zero $k\beta_0$} & \textbf{No KV} & \textbf{Wire} & \textbf{Loss / BW} & \textbf{No SM} \\
 & \textbf{Tax} & \textbf{Xfer} & \textbf{Preempt} & \textbf{Adaptive} & \textbf{Partitioning} \\
\midrule
Sarathi-Serve~\cite{agrawal2024sarathi} & $\times$ & $\checkmark$ & $\times$ & $\times$ & $\checkmark$ \\
DistServe~\cite{zhong2024distserve} & $\checkmark$ & $\times$ & $\times$ & $\times$ & $\checkmark$ \\
MuxWise~\cite{chen2026muxwise} & $\checkmark$ & $\checkmark$ & $\times$ & $\times$ & $\times$ \\
\revisit~\cite{hwang2026revisitpp} & $\times$ & $\checkmark$ & $\times$ & $\times$ & $\checkmark$ \\
\textbf{\sys} & $\checkmark$ & $\checkmark$ & $\checkmark$ & $\checkmark$ & $\checkmark$ \\
\bottomrule
\end{tabular}
\end{table}

Orca~\cite{yu2022orca} introduced iteration-level scheduling, and vLLM~\cite{kwon2023pagedattention} introduced PagedAttention. Sarathi-Serve~\cite{agrawal2024sarathi} batches prompt chunks with decode tokens. QoServe~\cite{goel2026qoserve} adjusts chunk size using request-level latency slack. \revisit~\cite{hwang2026revisitpp} tunes token chunk size via RLS and delays selected decodes on PCIe GPUs. AlpaServe~\cite{li2023alpaserve} optimizes model placement for statistical multiplexing, and ExeGPT~\cite{oh2024exegpt} schedules resources under inference latency constraints.

DistServe~\cite{zhong2024distserve}, Splitwise~\cite{patel2024splitwise}, and Mooncake~\cite{qin2025mooncake} serve prefill and decode on separate instances and transfer KV state between them. Llumnix~\cite{sun2024llumnix} migrates live requests. LoongServe~\cite{wu2024loongserve} adapts sequence parallelism as context lengths change, and Libra~\cite{ruan2026libra} partitions requests at flexible token positions. NanoFlow~\cite{zhu2025nanoflow} overlaps operations from different microbatches on a GPU. POD-Attention~\cite{kamath2025podattention} overlaps prefill and decode inside an attention kernel. MuxWise~\cite{chen2026muxwise} partitions GPU SMs between prefill and decode and schedules prefill layer by layer while sharing the KV cache. Table~\ref{tab:related} compares how four of these systems schedule computation and move data. In \sys, prefill and decode can each use the full stage GPU. Prefill yields at a layer boundary and bulk transfers yield at a frame boundary.

Petals~\cite{borzunov2023petals}, HexGen~\cite{jiang2024hexgen}, and PlanetServe~\cite{fang2026planetserve} distribute inference across heterogeneous or decentralized nodes. M\'elange~\cite{griggs2024melange} matches workloads to GPU types. PowerInfer~\cite{song2024powerinfer} splits neuron computation between CPU and GPU, and MegaScale-Infer~\cite{zhu2025megascaleinfer} disaggregates attention and expert computation.

SmoothQuant~\cite{xiao2023smoothquant} and QuaRot~\cite{ashkboos2024quarot} enable low-bit inference. CacheGen~\cite{liu2024cachegen} compresses KV state for transfer, and CacheBlend~\cite{yao2025cacheblend} reuses cached KV states for text chunks with selective recomputation. \sys compresses boundary activations so it can send them in small groups across pipeline stages.

%-------------------------------------------------------------------------------
\section{Conclusion}
\label{sec:conclusion}
%-------------------------------------------------------------------------------

\sys schedules wide prefill batches in GPU layer slices and transmits compressed activations in independently sized row groups. Decode work can preempt at layer boundaries, and the scheduler limits the time each row group occupies the link. In our evaluation, \sys delivers $23.3\times$ the SLO goodput of calibrated \revisit{} in four-stage MoE serving and reduces short-request p95 TTFT from $129.8$~s with native vLLM to $512$~ms in a mixed 64K-context workload.

%-------------------------------------------------------------------------------
\section*{Availability}
%-------------------------------------------------------------------------------
All source code for \sys, the fused Triton \hintfour kernels, the in-engine \revisit baseline port, per-request JSON replay logs, and automated reproduction scripts will be released as an open-source artifact upon publication.

\bibliographystyle{plainurl}
\bibliography{references}

\end{document}